\documentclass[12pt, letterpaper]{article}

\usepackage{amsmath,amssymb}
\usepackage{cite}
\usepackage{fancyhdr}
\usepackage[top=1in, bottom=1in, left=1in, right=1in]{geometry}
\usepackage[dvipdfm, dvips]{graphicx}
\usepackage{hyperref}

\numberwithin{equation}{section}

\begin{document}


\setcounter{page}{0}
\date{}

\lhead{}\chead{}\rhead{\footnotesize{RUNHETC-2012-07\\SCIPP-12/06}}\lfoot{}\cfoot{}\rfoot{}

\title{\textbf{Update on the Pyramid Scheme\vspace{0.4cm}}}

\author{Tom Banks$^{1,2}$ and T.J. Torres$^{2}$\vspace{0.7cm}\\
{\normalsize{$^1$NHETC and Department of Physics and Astronomy, Rutgers University,}}\\
{\normalsize{Piscataway, NJ 08854-8019, USA}}\vspace{0.2cm}\\
{\normalsize{$^2$SCIPP and Department of Physics, University of California,}}\\
{\normalsize{Santa Cruz, CA 95064-1077, USA}}}

\maketitle
\thispagestyle{fancy}

\begin{abstract}
\normalsize \noindent
We summarize recent work in which we attempt to make a consistent model of LHC physics, from the Pyramid Scheme.  The models share much with the NMSSM, in particular, enhanced tree level contributions to the Higgs mass and a preference for small $\tan\beta$.  There are $3$ different singlet fields, and a new strongly coupled gauge theory, so the constraints of perturbative unification are quite different.  We outline our general approach to the model, which contains a Kahler potential for three of the low energy fields, which is hard to calculate.   Detailed calculations, based on approximations to the Kahler potential, will be presented in a future publication. 
\end{abstract}


\newpage
\tableofcontents
\vspace{1cm}


\section{Introduction: The Pyramid Scheme}

The Pyramid Scheme is {\it not} the NMSSM, but has certain features which resemble it.  The Pyramid Scheme is an effective field theory, with a small number of parameters, which combines phenomenological constraints with theoretical constraints coming from the formalism of Holographic Space-time\cite{hst}.  In that formalism, the cosmological constant (c.c.) is a tunable parameter and SUSY is automatically restored in the limit of vanishing c.c., according to the formula
$$m_{3/2} = K \sqrt{\frac{m_P}{M_U}} \Lambda^{1/4} ,$$ where $M_U$ is the scale of standard model coupling unification, here identified with the inverse of the linear size of a roughly isotropic higher dimensional space.  $K$ is a constant which is roughly order $1$.

The limiting theory for $\Lambda = 0$ is super-Poincare invariant, and experience has shown that super-Poincare invariant models of quantum gravity obey the naturalness constraints of effective field theory.  All small numbers are explained by ratios of dynamical scales, or by approximate symmetries.  Approximate symmetries are rife in extreme limits of moduli space, but in HST the moduli are discrete for non-zero c.c. and cannot be too large.  

In effective field theory, the $\Lambda = 0$ theory should therefore possess an exact discrete $R$ symmetry, to explain the vanishing of the c.c. .  It was argued in \cite{susyhor} that interactions with the huge set of quantum states on the cosmological horizon of de Sitter (dS) space, generate R violating interactions, which in turn give rise to spontaneous SUSY breaking in a dS minimum of the effective field theory.  The R breaking constant $W_0$ in the effective field theory is tuned to obtain the value of the c.c. prescribed by the underlying HST model, which is NOT a quantum field theory.  All the R violating terms come from special diagrams, in which a single gravitino is exchanged with the horizon, and do not satisfy the naturalness constraints of QFT.  They must be computed from the underlying theory.

We do not know how to do these computations in detail.  However, we know that the effective field theory must have a dS minimum, with the c.c. tuned to the right value and must break R symmetry explicitly.  We also know \cite{tunneling} that this minimum must be absolutely stable when gravity is neglected.  In other words, the effective Lagrangian {\it must} be non-generic, in the language of Nelson and Seiberg\cite{ns}.   Apart from that, we use phenomenological constraints to fix the structure of ${\cal L}_{eff}$.  

SUSY is broken by the $F$ term of some low energy chiral field.  The size of this $F$ term is roughly
$$ F = K m_P  \sqrt{\frac{m_P}{M_U}} \Lambda^{1/4}. $$ For a unification scale $2 \times 10^{16} $ GeV, this is
$$F = 20K \sqrt{10} 10^6 ({\rm GeV})^2 \sim (8\sqrt{K} {\rm TeV} )^2 .$$  In order to generate gaugino masses, the chiral field must have a dimension $5$ coupling to the gauge kinetic terms, leading to

$$m_{1/2}^i = X_i \frac{\alpha_i}{4\pi} \frac{F}{\Lambda_3} .$$  This coupling doesn't vanish in the limit of vanishing c.c., and must obey the constraints of naturalness.  Thus, given the small size of $F$, the scale $\Lambda_3$ must be the dynamical scale of a new strongly coupled gauge theory. Furthermore, there must be fields in this theory that are charged under all components of the standard model gauge group.

We now have a potential challenge for standard model gauge coupling unification.  To preserve it, the new charged fields must be in complete multiplets of the unified gauge group\footnote{However, see the comments about accelerated unification in the conclusions.}, and the number of new multiplets must not be too large.  We also require that the model have a dark matter candidate, and no pseudo-Nambu Goldstone bosons with standard model quantum numbers (which are experimentally ruled out for the values of the relevant scales).  Once two loop renormalization group corrections are taken into account, the only models which appear to survive are the Pyramid Schemes.

In the Pyramid Schemes, unification is Trinification\cite{trinification}.  The new strong coupling gauge group is $SU_P (k)$ with $k=3,4$ and the new matter fields, called {\it trianons} , $T_i \oplus \tilde{T}_i$, are in the $(k,\bar{3_i}) + (\bar{k} , 3_i ) $ of $SU_P (k) \times SU_i (3)$, where the latter factor is the $i$th factor of the trinification group $SU(3)^3 \ltimes Z_3$.  The quivering moose diagram of this model is a three sided pyramid.  The base of the pyramid is a triangle composed of 3 generations of chiral fields in the $(1,1,\bar{3}, 3) + (1,3,1,\bar{3}) + (1,\bar{3},3,1)$ of $SU_P (k) \times [SU(3)^3 \ltimes Z_3 ]$.  The trianons $T_i$ and $\tilde{T}_i$  are cyclically permuted by $Z_3$.  

The trinification group is broken at the unification scale $M_U = 2 \times 10^{16}$ GeV, leaving over only $15$ states from each generation.  We do not specify where the Higgs fields $H_{d,u}$ arise at the unification scale. All questions of the Yukawa textures of the Higgs coupling in generation space, as well as the origin of the neutrino masses (and the discrepancy of more than an order of magnitude between $M_U$ and the mass parameter $M_{\nu}$ in the dimension $5$ operator $\frac{1}{M_{\nu}} (L H_u)^2$), are considered unification scale physics and will not be discussed here.  All we ask of our low energy model is that it has a particle content and couplings consistent with unification.

Most of the work on this model has been done for $k=3$, and we will only describe that case here.  Below the confinement scale $\Lambda_3$ of the Pyramid gauge group there are standard model singlets that are baryons of $SU_P (3)$.  The gauge interactions preserve all three trianon numbers, but the renormalization group structure of the model requires at least two of them to be broken\cite{pole}.  The reason is that $SU_P (3)$ is not asymptotically free at high energies, and only becomes so below the scale of the highest trianon mass. However, it is well known\cite{evashamitleighstrassler} that if we add trilinear superpotential terms for all of the trianons, with couplings equal to the $SU_P (3)$ gauge coupling, then we have a line of fixed points.  In \cite{pole} it was shown that even with only two of the trilinear terms, the gauge coupling can hover near a large value between the scales $10^{16}$ GeV and $1$ TeV.  This allows the confinement scale $\Lambda_3$ to be close to the trianon masses, in a model which has a conserved trianon number for one of the trianons.  The lightest bound state carrying this quantum number is called the {\it pyrma-baryon} and is the dark matter candidate of the Pyramid Scheme.

We choose either of $T_{1,2}$ to be the conserved particle number, as a consequence of which the {\it pyrma-baryon} has a magnetic dipole moment.  This gives rise to interesting direct detection signals\cite{bft} and may have interesting astrophysical consequences.  The cosmologically stable pyrma-baryon can be the dark matter if the right asymmetry is generated in the early universe.  There is a (roughly calculable) dimension six interaction between this conserved current and ordinary baryon and lepton numbers\cite{bhbej}, which leads to several possible scenarios for connecting the dark matter density to the baryon asymmetry.   

The trianons all have mass, and we assume these mass terms are R violating operators, coming from interactions with the horizon.  In order to use Seiberg's solution of SUSY QCD with $N_F = N_C$ we assume two of these masses are slightly above the confinement scale, and one sufficiently far below that we can use Seiberg's effective Lagrangian on moduli space.  
In terms of familiar QCD scales, we can think of two of the trianon mass to confinement scale ratios as of order the analogous ratio for the $\rho$ meson , while the colored trianon has mass to confinement scale ratio of order that of the strange quark.
We assume the colored trianon is the light one because of the experimental bound on gluino masses.  If we took the colored trianon mass to be large, the gluino mass would go to zero.  

Another way to state this assumption about the masses is simply to say that two of them are a factor of $5$ or so larger than the third, and that the $SU_P (3)$ gauge coupling is at the edge of the perturbative regime at the scale of the heavier masses.  This means that the confinement scale is close to but below the heavier masses, while the third mass is light enough to use chiral perturbation theory.

The model as described thus far does not break SUSY, and so cannot be the effective theory of an underlying HST model of dS space.  To break SUSY, we add singlets.   It turns out that three singlets is the minimum to make a working model.  We assume R preserving trilinear couplings of the singlets to each of the $T_i \tilde{T}_i$ bilinears, as well as to $H_u H_d$.  The R symmetry can be chosen to allow all these couplings, forbid all the corresponding bilinears, and also forbid all B and L violating operators of dimensions $4$ and $5$\cite{pyramid1}.  R violation gives us the bilinears, and we assume it also gives linear terms in the singlets, which will give rise to SUSY breaking.  Above the confinement scale $\Lambda_3$, the effective superpotential is

$$ W = \sum_i (m_i + \alpha_i^j S_j) T_i \tilde{T}_i  + (\mu + \beta^i S_i) H_u H_d   + \sum _i (g_i T_i^3 + \tilde{g_i} \tilde{T}_i^3) + F^i S_i + W_{std}.$$ The last term is the standard model superpotential.  In this equation $T_i^3$ is shorthand for the invariant $\epsilon_{abc} \epsilon^{ABC} (T_i )_A^a (T_i )_B^b (T_i )_C^c$, where the small indices are triplets of $SU_P (3)$ and the capital indices anti-triplets of the $SU_i (3)$ subgroup of the trinification group.
We also have $g_i = \tilde{g}_i = 0$ for either $i = 1$ or $2$.  We integrate out the colorless trianons, and use Seiberg's Lagrangian for  $N_F = N_C = 3$ QCD.   We'll also assume the vacuum preserves color.  
With these assumptions, the low energy effective superpotential is
$$ W_{eff} = 3 (m_3 + \alpha_3^j S_j) M \Lambda_3 + (\mu + \beta^i S_i) H_u H_d + F_i S_i + g_3 B \Lambda_3^2 + \tilde{g}_3 \tilde{B}\Lambda_3^2 
+ L( M^3 - \Lambda_3 B\tilde{B} - \Lambda_3^3).$$ $M$ is the coefficient of $1$ in the meson matrix, $M_A^B = (T_3 )_A^a (\tilde{T}_3 )^B_a$, and we've taken the non-dynamical Lagrange multiplier field $L$ to be dimensionless.  In principle, we can add cubic and quadratic terms in the singlet fields, as long as they preserve the property, which we will now demonstrate, that the superpotential has no stationary points, so that SUSY is broken.  The parameters $m_i, F_i$ and $\mu$ arise from R violating interactions with the horizon, and vanish with the c.c., while the others are independent of the c.c. when it is small.

In writing this superpotential, we are implicitly making the assumption that the value of the singlet fields that minimizes the potential is such that the masses of the colorless trianons are larger than the scale $\Lambda_3$.  We will assume that they are not enormously larger, since in that limit the $SU(2) \times U(1)$ gaugino masses go to zero. This phenomenological constraint fits with our theoretical expectations.  The parameters $m_i$ are determined by the same sort of gravitino exchange diagrams that fix the SUSY breaking parameters $F_i$.

The F term equations for the fields $L,B,\tilde{B}$ fix these fields in terms of $M$ and leave over an $M$ dependent superpotential
$$g B + \tilde{g} \tilde{B} = 2\sqrt{g\tilde{g}(M^3 -1)}.$$  
The F term equations for the singlets read
$$F^i  +3 \alpha_3^i M + \beta^i H_u H_d  = 0.$$  If the $3-$ vectors $F^i$, $\alpha_3^i $ and $\beta^i$ are linearly independent, these are $3$ equations for two unknowns, and have no solution. SUSY is broken and the non-vanishing F term at the minimum is of order the coefficients $F^i$.  

If the Kahler potential were canonical, we could choose a basis in field space where the $F$ term corresponds to a single decoupled chiral multiplet.  However, the Kahler potential is a non-trivial function of $M$ and of the combinations $\alpha_a^j S_j$ with $a = 1,2$.  In principle it also depends on {\it e.g.} $B + \tilde{B}$, but we have enough equations to eliminate these fields in favor of $M$.  The scales in the Kahler potential are $m_a$ and $\Lambda_3$, which are roughly comparable.
Hard core effective field theorists will have noticed several issues of ``tuning" in the above discussion.  We will return to them in the conclusions.  

We've seen that for $\Lambda_3 \sim 1 $ TeV and $F$ given by the HST relation between the gravitino mass, the c.c. and a unification scale of order $2 \times 10^{16} $ GeV, we get acceptable values for the gaugino masses.
For this value of $\Lambda_3$ it is also plausible that $SU_P (3)$ ``hadrons" with standard model quantum numbers will be compatible with the fairly stringent experimental bounds on new colored states that have been established at the LHC.  

\subsection{The Higgs potential}

The tree level potential following from the superpotential and Kahler potential we have written has a competition between two effects.  The F terms of the singlet fields prefer a non-zero value for $H_u H_d$, whereas the $F$ terms of the Higgs fields themselves prefer the Higgs VEVs to vanish, if the effective $\mu$ parameter, $\mu_{eff} = \mu + \beta^i S_i$ is non-zero.   There are so many ways that the potential depends on the singlets, that it is hard to imagine that $\mu_{eff}$ will in fact vanish. Without a detailed knowledge of the Kahler potential we cannot check this.

We should also take into account the large radiative correction from the non-cancelation of top and stop loops, which gives rise to

$$\delta V_{eff} = - \frac{12 y_t^2}{16\pi^2} |H_u |^2 m_{\tilde{t}}^2 {\rm ln} (\Lambda_c / m_{\tilde{t}}) .$$
In a gauge mediated model like the Pyramid scheme we have
$$m_{\tilde{t}} = X \frac{\alpha_3 F}{4\pi\Lambda_3} \sim .6 X {\rm TeV} .$$
We'll argue in the conclusions that the cutoff scale should be of order at most$10$ TeV, so that the logarithm is $\sim 2$.  If the stop mass is $1$ TeV, then the effective negative mass squared parameter is about
$$m_{eff}^2 \sim \frac{24}{4\pi} \frac{m_t^2}{m_W^2} (1 + \cot^2 \beta)^2 ({\rm TeV})^2 .$$  This term is about the same size as other terms in the potential.  It is important, because without it the theory makes the prediction that $\tan\beta = 1$, which is inconsistent with perturbative unification at the conventional unification scale.  The top/stop loop correction favors larger $\tan\beta$ so the perturbative unification bound of $\tan\beta > 1.7$ or so is, plausibly, satisfied. 
Since most of the parameters in the Higgs potential are of order a TeV, it would appear that the Pyramid scheme has a bit of a little hierarchy problem.
It is hard to assess the extent of this tuning without a  calculation of the Kahler potential, though it is nominally of order $1\%$.  In a paper in preparation\cite{tbtjt}, where we have made some simple approximate evaluations of $K$, this is indeed the degree of tuning we find. We want to emphasize that in the Pyramid scheme this problem is independent of the stop mass.  The tree level potential contains parameters which are forced to be of order TeV scale, by matching to the scale of SUSY breaking in the underlying HST model.  These tend to make the Higgs VEV too large, unless there is some sort of cancelation. 

On the other hand, there is no apparent problem with obtaining a Higgs mass of order $125$ GeV.  We have several singlets coupled to $H_u H_d$ so there does not appear to be a problem with perturbative unification below the Landau pole for these couplings\footnote{Again, without a full evaluation of the Kahler potential near the minimum, it is hard to make these assertions with certainty, since the normalizations of the fields appearing in the superpotential are not canonical.}.

\section{Discussion and Conclusions}

Although we will not enter into the details here, the discrete R symmetry of the $\Lambda = 0$ model can be chosen\cite{pyramid1}\cite{pole} to eliminate all dimension $4$ and $5$ B and L violating operators in the supersymmetric standard model, with the exception of $H_u^2 L^2$.  Although the R symmetry is broken, the breaking goes via a special class of diagrams in which a pair of gravitinos is exchanged with the horizon. These diagrams obey a selection rule: R charge is violated by exactly two units.  In addition, we want to argue that the contribution to these diagrams from high energy physics is power law suppressed.

The diagrams are evaluated to leading order in small $\Lambda$.  In order to get a finite answer, we have to keep the $\Lambda$ dependence in the gravitino mass, but all the vertices in the part of the diagram localized near the observer at the origin of static coordinates are evaluated with $\Lambda = 0$.   Since the $\Lambda = 0$ theory has no $B$ violation at dimension $4$ and $5$, B violating operators will be of effective dimension $6$ or higher, even when R breaking is taken into account.

A similar argument gives a novel solution of the strong CP problem.  The low energy $\Lambda = 0$ theory has $U(1)$ axial symmetries, which can eliminate all CP violating phases besides the phase in the CKM matrix.
In effect, it solves the strong CP problem by having an axion-like field.
The R breaking operators give this axion a large mass, nominally of order $100$s of GeV to several TeV.  In ordinary effective field theory, we would argue that these R breaking terms come equipped with phases, which re-introduce the strong CP problem.  However, the diagrams that give rise to them involve only low energy interactions, which violate CP only through the CKM matrix, higher dimension operators suppressed by the unification scale, 
and the interactions of the gravitino with the dS horizon.  The latter interaction takes place at a very high temperature, of order the Planck scale, so if the fundamental origin of CP violation is at a scale below the Planck scale, there will be no phase in the gravitino-horizon interactions\footnote{We note that these interactions involve second order perturbation theory, so it may be that phases cancel even if there is no energy suppression of CP violation.}.   $\theta_{QCD}$ in the Pyramid scheme appears to be dominated by the standard model loop corrections involving the CKM phase, and this value is far below the experimental bounds.

The peculiar nature of the R violating interactions in the Pyramid scheme also accounts for the low value of the cutoff in the calculation of the stop loop contribution to the Higgs potential.  The values of these operators at the TeV scale are determined by consistency with the scale of SUSY breaking dictated by the underlying HST model and the observed value of the cosmological constant .
At higher scale these diagrams are power law suppressed because they require two very long wavelength gravitino lines to couple to an operator of very small space-time extent.  

In conclusion, the Pyramid scheme, with the value of the SUSY breaking F term indicated by HST and the observed c.c., can give a satisfactory account of current experimental data, but requires a fine tuning, which is nominally of order $1\%$ to explain the value of the standard model Higgs VEV in terms of parameters of the TeV scale.  In this model, even a light stop does not improve the situation.  The precise degree of fine tuning, as well as the detailed spectrum of new particles, depends on a Kahler potential which we can only estimate at present.

The dark matter candidate of the model is a hidden sector baryon with a mass in the multiple TeV range, and a magnetic moment.  The model preserves baryon and lepton numbers at the dimension $\leq 5$ level, and resolves the strong CP problem, because of novel selection rules for interactions induced by degrees of freedom on the cosmological horizon.
These interactions break a discrete R symmetry explicitly, and lead to SUSY violation in low energy effective field theory, without R symmetry or an R axion.  They avoid the Nelson-Seiberg theorem\cite{ns} because they do not satisfy the criteria of naturalness.  They are not the most general interactions consistent with unbroken symmetries.  Calculations in perturbative string theory lead us to expect field theoretic naturalness criteria to be correct in models of quantum gravity in asymptotically flat space.
However, interactions of localized matter with the cosmological horizon are mediated by very special two gravitino exchange diagrams and violate field theoretic naturalness.

Like many models with extra singlets, the Pyramid scheme has no trouble generating a Higgs mass of order $125$ GeV, once the tuning of the electroweak scale is performed.  We note that interactions with the horizon generate a $\mu$ term, independently of the singlet VEVs, so some of the conventional constraints on models with extra singlets are avoided.  Although a proper calculation awaits better understanding of the Kahler potential, the fact that there are $3$ different singlets could also ease constraints coming from requiring that the singlets remain weakly coupled up to the unification scale.

We also note that, since the trinification scheme does not have B violating gauge interactions, it might be easier to lower the unification scale\cite{acceluni} without running into problems with proton decay.  Note however that in HST, lowering the unification scale leads to an increase in the SUSY breaking scale, which would appear to make the hierarchy between the electroweak scale and the SUSY scale more problematic.

Finally, we want to comment on the coincidence between the SUSY breaking scale and the confinement scale $\Lambda_3$.  There is a curious mix of environmental selection and dynamics in this condition.  In HST, the coupling of the $SU_P (3)$ gauge theory at the unification scale is well approximated by its value in a model with zero c.c. .  In that limit we have an isolated super-Poincare invariant model with a discrete R symmetry. There are no known models of this type, even in the hypothetical ``string theory Landscape".  It is natural to assume a certain degree of uniqueness for such models, and to assume in particular that the $SU_P (3)$ coupling is completely fixed.

The R breaking masses for the trianon fields come from interaction with the cosmological horizon, and depend on the c.c. .  In HST cosmology the universe is a distribution of (future) asymptotically dS island universes, embedded in a $p= \rho$ background space-time.  From the point of view of the background, the dS islands look like black holes of varying Schwarzschild radii.  Depending on multiversal initial conditions\footnote{The relative positions and velocities of the black holes in the background space-time.}, some of them may collide with each other and the time scale between such collisions appears to be tunable.  The distribution of cosmological constants makes this model a candidate for environmental selection.

When all trianons are massless, the $SU_P (3)$ gauge coupling is not asymptotically free, but as a consequence of two trilinear terms in the superpotential, one is close to the fixed line behavior of the quiver gauge theory with $9$ flavors\cite{evashamitleighstrassler}.  In \cite{pole}, the authors argued that the couplings would hover near a fixed line (the position on that line determined by values at the unification scale), and then slowly asymptote to zero.  Thus, if we choose a small c.c., which implies small breaking of R symmetry and SUSY, the trianons have very small masses and $SU_P (3)$ is very weakly coupled.  The trianons must be included in the low energy spectrum, and QCD is not asymptotically free. In addition, the world is close to being exactly supersymmetric.  Hadrons, atoms, stars {\it etc.} don't exist in this world and the parameter $\Lambda_3$ is not defined.

Now consider what happens when the c.c. is chosen much larger than the value required to fit the real world.   The trianons are now very heavy and can be integrated out.  SUSY is broken at a relatively high scale and we can use the constant in the superpotential to fit the HST relation between the c.c. and the gravitino mass.  However, the gauge mediated contributions to SUSY breaking in the standard model are highly suppressed.

The couplings of Higgs fields to the singlets are unchanged, so even if we allow a one percent fine tuning, the breaking of electroweak symmetry will occur at a scale much higher than it does in the real world, and quark masses will be rescaled as well. Generically, as in the phenomenologically acceptable Pyramid Scheme, there will be F terms for the Higgs superfields.  Since SUSY breaking among standard model multiplets is dominantly mediated by their coupling to the Higgs, the light quark, and all lepton, multiplets will be very close to supersymmetric. There will be no stable nuclei or atoms.

Thus, allowing only the c.c. to be scanned as we look at different isolated asymptotically dS universes in the HST multiverse, very few of these island universes will have low energy nuclear and atomic physics that resembles our own.   Thus, an island universe with a c.c. such that there is a confinement scale $\Lambda_3$ and (for self consistency) the trianon mass terms are close to this scale, will be environmentally selected by atomic and nuclear physics.  The degree of precision with which this criterion fixes the c.c. to its observed value is hard to compute, but it is clear that it cannot vary by more than an order of magnitude or two.

It should be emphasized that this anthropic argument does not guarantee the {\it orderings} of trianon masses, which we have found necessary for more detailed reproduction of experiment.  This must be justified by detailed calculations of particle interactions in the underlying HST model.  At the present time we are not able to do those calculations.  

Another important point to emphasize is that these considerations fix the c.c.
without invoking Weinberg's \cite{weinberg} galaxy formation bound.  That bound should be thought of as a lower bound on the product of the dark matter density and the primordial fluctuation amplitude, $Q$.  In the Pyramid Scheme, dark matter is a baryon, $T_a^3$, with $a = 1$ or $2$.  The dark matter density is determined by physics at the unification scale, which produces an asymmetry.  It seems unlikely that this is affected by scanning the c.c..   On the other hand, in the HST model of inflationary fluctuations\cite{holoinflation}, $Q$ definitely takes on different values in different island universes. Galaxy formation should be thought of as an environmental selection criterion for $Q$, with the c.c. and dark matter densities fixed by other criteria.

The Pyramid scheme, for values of its parameters which are generic within the framework of its assumptions, gives a consistent explanation of all extant data and a wealth of new predictions about supersymmetry, and the nature of dark matter.   Precise predictions for the particle spectrum require the solution of a strongly coupled supersymmetric gauge theory, including the calculation of the Kahler potential on its moduli space.  One clear prediction is that the vacuum angle satisfies $\tan\beta\sim 1$, with the deviation from $1$ driven by the top/stop contribution to the Higgs potential. There appears to be a little hierarchy problem: a tuning of order $1\%$ is required in order to explain the electroweak VEV of order $174$ GeV in terms of parameters that are of order $1$ TeV.  There does not appear to be a $\mu$ or $B_{\mu}$ problem, because the model incorporates singlets, as in the NMSSM.  

The Pyramid Schemes present a novel solution of the strong CP problem. The $\Lambda =0$ effective Lagrangian, ${\cal L}_0$ has axion-like $U(1)$ symmetries, which allow us to rotate away all CP violating phases. The corrections to ${\cal L}_0$ come from special diagrams where a gravitino is exchanged with the horizon, which lift the would be axion masses to the TeV scale, without introducing new phases.  A similar mechanism is used in the explanation of the absence of baryon number violating operators of dimension $4,5$.

\vspace{2pc}

\begin{center}
{\bf  Acknowledgements:}
\end{center}
 This work was supported in part by the U.S. Department of
Energy.\vfill\eject



\begin{thebibliography}{99}
\bibitem{hst} T.~Banks,
  ``Holographic Space-Time: The Takeaway,''
  arXiv:1109.2435 [hep-th];
   T.~Banks,
  ``TASI Lectures on Holographic Space-Time, SUSY and Gravitational Effective Field Theory,''
  arXiv:1007.4001 [hep-th];
  T.~Banks,
  ``Holographic space-time and its phenomenological implications,''
  Int.\ J.\ Mod.\ Phys.\ A {\bf 25}, 4875 (2010)
  [arXiv:1004.2736 [hep-th]];
 T.~Banks,
  ``Deriving particle physics from quantum gravity: A Plan,''
  arXiv:0909.3223 [hep-th];
   T.~Banks,
  ``Holographic Space-time from the Big Bang to the de Sitter era,''
  J.\ Phys.\ A A {\bf 42}, 304002 (2009)
  [arXiv:0809.3951 [hep-th]].

\bibitem{pyramid1}  T.~Banks and J.~-F.~Fortin,
  ``A Pyramid Scheme for Particle Physics,''
  JHEP {\bf 0907}, 046 (2009)
  [arXiv:0901.3578 [hep-ph]].
\bibitem{trinification}  S.~L.~Glashow,
  ``Trinification Of All Elementary Particle Forces,''
  Print-84-0577 (BOSTON).



\bibitem{susyhor}  T.~Banks,
  ``SUSY and the holographic screens,''
  hep-th/0305163.

\bibitem{tunneling}
  T.~Banks and J.~-F.~Fortin,
  ``Tunneling Constraints on Effective Theories of Stable de Sitter Space,''
  Phys.\ Rev.\ D {\bf 80}, 075002 (2009)
  [arXiv:0906.3714 [hep-th]].

\bibitem{ns}  A.~E.~Nelson and N.~Seiberg,
  ``R symmetry breaking versus supersymmetry breaking,''
  Nucl.\ Phys.\ B {\bf 416}, 46 (1994)
  [hep-ph/9309299].

\bibitem{pole}T.~Banks, J.~-F.~Fortin and S.~Kathrein,
  ``Landau pole in the pyramid scheme,''
  Phys.\ Rev.\ D {\bf 82}, 115015 (2010)
  [arXiv:0912.1313 [hep-ph]].


\bibitem{bhbej} T.~Banks and H.~E.~Haber,
  ``Note on the pseudo-Nambu-Goldstone Boson of Meta-stable SUSY Violation,''
  JHEP {\bf 0911}, 097 (2009)
  [arXiv:0908.2004 [hep-ph]];
 T.~Banks, S.~Echols and J.~L.~Jones,
  ``Baryogenesis, dark matter and the Pentagon,''
  JHEP {\bf 0611}, 046 (2006)
  [hep-ph/0608104];
 T.~Banks, J.~D.~Mason and D.~O'Neil,
  ``A Dark matter candidate with new strong interactions,''
  Phys.\ Rev.\ D {\bf 72}, 043530 (2005)
  [hep-ph/0506015].

\bibitem{tbtjt} T.~Banks, T.J.~Torres, {\it Approximate Particle Spectra in the Pyramid Scheme}, paper in preparation.
\bibitem{evashamitleighstrassler}   R.~G.~Leigh and M.~J.~Strassler,
  ``Exactly marginal operators and duality in four-dimensional N=1 supersymmetric gauge theory,''
  Nucl.\ Phys.\ B {\bf 447}, 95 (1995)
  [hep-th/9503121].

S.~Kachru and E.~Silverstein,
  ``4-D conformal theories and strings on orbifolds,''
  Phys.\ Rev.\ Lett.\  {\bf 80}, 4855 (1998)
  [hep-th/9802183].
\bibitem{bft} T.~Banks, J.~-F.~Fortin and S.~Thomas,
  ``Direct Detection of Dark Matter Electromagnetic Dipole Moments,''
  arXiv:1007.5515 [hep-ph].
\bibitem{acceluni} N.~Arkani-Hamed, A.~G.~Cohen and H.~Georgi,
  ``Accelerated unification,''
  hep-th/0108089.

\bibitem{weinberg} S.~Weinberg,
  ``Anthropic Bound on the Cosmological Constant,''
  Phys.\ Rev.\ Lett.\  {\bf 59}, 2607 (1987).

\bibitem{holoinflation}T.~Banks and W.~Fischler,
  ``Holographic Theories of Inflation and Fluctuations,''
  arXiv:1111.4948 [hep-th].

\end{thebibliography}


\end{document}